\documentclass[12pt]{article}
\usepackage{pdproc,epsfig}
\newcommand{\be}{\begin{equation}}
\newcommand{\ee}{\end{equation}}
\newcommand{\bea}{\begin{eqnarray}}
\newcommand{\eea}{\end{eqnarray}}
\newcommand{\bi}{\bibitem}

\newcommand{\rar}{\rightarrow}
\newcommand{\lrar}{\leftrightarrow}

\newcommand{\mnu}{m_\nu}

\newcommand{\hf}{\hat f}
\newcommand{\hb}{\hat b}
\newcommand{\hfc}{\hat f^+}
\newcommand{\hbc}{\hat b^+}
\newcommand{\cm}{c_1}
\newcommand{\sm}{s_1}

\begin{document}

\renewcommand{\thefootnote}{\alph{footnote}}
  
\title{
NEUTRINO, COSMOS, AND NEW PHYSICS
}

\author{ A.D. DOLGOV}

\address{ 
INFN, Ferrara 40100, Italy\\
ITEP, 117218, Moscow, Russia\\
 {\rm E-mail: dolgov@fe.infn.it}}

\abstract{
Observational manifestations of possible breaking of spin-statistics
relation for neutrinos are considered. It is argued that bosonic 
neutrinos may form cosmological cold dark matter, improve agreement 
of BBN predictions with observations, make operative Z-burst model of
ultra-high energy cosmic rays, etc. Restrictions for an admixture of
bosonic component to neutrino which follow from double beta decay are
discussed.
}
   
\normalsize\baselineskip=15pt

\section{Introduction \label{s-intr} }
  
There is an impressive symbiosis of the ``weakest'' and lightest of the known 
massive particles, neutrino, and Cosmos. Cosmology and astrophysics
allow to study neutrino properties with an accuracy which, in many cases,
is unaccessible in direct terrestrial experiments and, vice versa, neutrino
helps to resolve some cosmological and astrophysical 
mysteries\cite{nu-rev,han-rev}. 

Cosmology allows to put a very stringent upper bound on neutrino mass 
at the level of about 1 eV, for a review see e.g. papers\cite{nu-mass}. 
The bound on the amplitude of possible right handed currents and the
mass of right intermediate bosons found from the analysis of Big Bang
Nucleosynthesis (BBN) is orders of magnitude better than those obtained 
in laboratories. The same is true for neutrino magnetic moments and 
mixing of the usual active neutrinos with hypothetical sterile
ones which are restricted by BBN\cite{nu-rev} and stellar 
evolution\cite{nu-astro}. 

On the other hand, neutrino contributes to cosmology providing hot dark
matter, but not the necessary cold one, if physics is normal. Neutrino may
be related to dark energy\cite{en-dark}, and be responsible, at least 
partly, for ultra-high energy cosmic rays beyond GZK cut-off through 
the $Z$-burst mechanism\cite{z-burst}. The large mixing angle solution
to the solar neutrino anomaly excludes noticeable lepton asymmetry of the
universe\cite{lept-asym}. 

These lists are far from being complete but this is not the main subject 
of this talk. Instead of these 
rather well known topics I would like to talk about new,
though quite speculative issues, related to effects of possible
breaking of neutrino statistics in cosmology. The content of the
talk is strongly based on ref.\cite{ad-as}.

The first question is why neutrino? First of all, neutrino is the only
known particle indicating to new physics. As is known from observation of
neutrino flavor oscillations leptonic flavor charges, electronic, muonic,
and tauonic, are not conserved. Neutrino is the only observed particle
for which Majorana mass is possible and, as a result total leptonic charge
could be non-conserved. There are experimental indications to 
a possible leptonic charge non-conservation from
neutrinoless double beta decay\cite{nu-less}. It may mean that neutrino
communicates with a hidden sector of the particle world and is a messenger
of new physics from the hidden sector. It could be that no sacred principles
are respected in the hidden world and neutrinos brings us exotic 
possibilities of breaking CPT theorem\cite{cpt} or Lorentz 
invariance, which are actively discussed in the recent years.
Since at the present days cosmology is quickly becoming precise science,
maybe cosmos will bring through neutrinos new surprising physics.

The most exciting possibility which, in particular, may lead to violation of 
CPT and Lorentz invariance and even to much more drastic consequences 
is a breaking of the spin-statistics 
relation for neutrinos. Ironically the particle brought to this world by Pauli
may violate Pauli exclusion principle. In fact Pauli and Fermi repeatedly
asked the question if spin-statistics relation could be not exact and 
electrons were a little bit different.

Possible violation of exclusion principle for the usual matter, i.e. for
electrons and nucleons was discussed in a number of papers at the end of
the 80th\cite{en-viol}. Efforts to find a  
more general than pure Fermi-Dirac or Bose-Einstein 
statistics\cite{para} were taken but no satisfactory theoretical 
frameworks had been found. Experimental searches of the
Pauli principle violation for electrons~\cite{exp-viol}  
and nucleons~\cite{exp-bar} have also given negative results.  

If one assumes that spin-statistics relation is broken while otherwise
remaining in the frameworks of the traditional quantum field theory then
immediately several deep theoretical problems would emerge:\\
1) non-locality;\\
2) faster-than-light signals;\\
3) non-positive energy density and possibly unstable vacuum;\\
4) maybe breaking of unitarity;\\
5) broken CPT and Lorentz invariance (as mentioned above).\\

Either these consequences (if they indeed were realized) exclude
any violation of spin-statistics
theorem and discussion of this violation should be forbidden or they
open an exciting space for further research and development.
An answer to that is first of
all a matter of experiment which may either exclude or confirm the
drastic assumption of breaking the spin-statistics relation. 
As for observational manifestations of the mentioned
phenomena they should be weak because they are 
induced by weakly interacting neutrinos and, moreover, in higher orders 
of perturbation theory.

Perturbative expansion of the scattering matrix has the well known form:
\be
S = 1 + \sum_n \frac {(-i)^n}{n!} \int \Pi d^4 x_j 
T\left\{ {\cal H}(x_1)... {\cal H}(x_n)\right\} 
\label{S}
\ee
where $T\{...\}$ means time-ordered product of operators inside brackets.
Lorentz invariance is ensured if Hamiltonian, $ {\cal H}$, is bosonic 
operator, i.e. it commutes with itself if separated by the light cone, see 
e.g.\cite{it-zub}. However, for bosonic or partly bosonic neutrinos the
effective Hamiltonian responsible even for the simple
reaction $e + p \leftrightarrow n+ \nu$ is
not bosonic and observables do not commute for space-like separation and 
locality breaks.
Presumably unitarity is maintained because Hamiltonian remains Hermitian.

Another possibility that Hamiltonian/Lagrangian approach and least
action principle are applicable only approximately and theory is drastically
modified, while the observable effects may still be small.

So let us postpone discussion of (non-existing) theory and consider
phenomenology of neutrinos obeying Bose or mixed statistics\cite{ad-as}.
What can we buy for this price?

\section{ Dark matter \label{s-dm}.}

It is well known that the usual fermionic neutrinos cannot form cosmological
cold dark matter for any spectrum of primordial fluctuations and arbitrary
self-interaction. This conclusion is based on the Tremain-Gunn 
bound\cite{tg} which does not allow to fill galaxies with sufficiently many 
light fermions (satisfying the Gerstein-Zeldovich bound\cite{gz})
to account for the observed hidden mass.
Thus we face the following dark matter dilemma: \\
1) new particles and old (normal) physics\\
2) old particles (neutrinos) and very new physics.

To make the cosmological cold dark matter neutrinos must form Bose 
condensate in the early universe. To this end a very large lepton asymmetry 
is necessary with
\be
\frac{|n_\nu - n_{\bar\nu}|}{n_\gamma} \sim 100
\label{n-nu-n-gamma}
\ee
Such asymmetry might be created in a version of the Affleck and 
Dine\cite{af-di} scenario.

A large asymmetry allows to fill up the present day universe by a huge
number of cosmic neutrinos such that they 
would be able to make all CDM, $\Omega_{CDM} \approx 0.25$, if
\be
n_\nu \sim 10^4   \,\, {\rm cm}^{-3}
\label{n-nu}
\ee
The spectrum of cosmic background neutrinos, if they are bosonic,
would be very much different from the fermionic ones
because the Bose equilibrium distribution has the form:
\be
f_{\nu_b} = \frac{1}{\exp [(E-\mu_\nu)/T -1 } + C \delta (k),
\label{f-b-nu}
\ee
where $\mu=m_\nu$ is the maximum value of chemical potential of bosonic 
neutrinos equal to their mass. The condensate amplitude
$C$ does not depend upon neutrino energy but may depend upon time.
One sees that the bulk of the condensed neutrinos is cold. 
In galaxies the neutrino number density would be about 6 orders of 
magnitude larger than the average cosmological number density, i.e. 
\be
n_\nu^{(gal)} \sim 10^{10}\left(m_\nu/0.1\,{\rm eV}\right){\rm cm}^{-3}.
\label{n-gal}
\ee

Structure formation with Bose condensed light bosons with the usual integer 
spin was considered in ref.\cite{mad}. The model well reproduces the 
essential features of the observed large scale structure. Since the picture
is spin independent the same must be true as well for bosonic neutrinos.

The results and numerical estimates presented in this section are true 
for purely bosonic neutrinos however, as we
see below, experiments on two neutrino double beta decay seem to exclude
100\% breaking of statistics and at least some fermionic fraction
must be present in a neutrino. It makes the model noticeably more
cumbersome, but less vulnerable.

\section{Equilibrium distribution for mixed statistics \label{s-equil}}

The statistics dependent term in kinetic equation for the reaction
$1+2\lrar 3+4$ has the form
\be
F = f_1(p_1) f_2(p_2) [1\pm f_3 (p_3)] [1\pm f_4 (p_4)]  \nonumber\\
- f_3(p_3) f_4(p_4) [1\pm f_1 (p_1)] [1\pm f_2 (p_2)]
\label{F}
\ee
where $f_j$ is the distribution function of particle $j$. This expression
is valid in the case of T-invariant theory when the detailed balance 
condition is fulfilled. Since T-invariance is broken, kinetic equation
is modified but the equilibrium distributions $f_j^{(eq)}$ remain the same
canonical Bose and Fermi ones as in T-invariant theory~\cite{ad-balance}. 
This statement is based on the unitarity of $S$-matrix. If the spin-statistics
relation is broken, as a result the unitarity may also be broken. If this
is the case, then a breaking of $T$-invariance may create large deviations
from the standard equilibrium distribution functions of neutrinos.

In what follows we neglect complications related to a 
violation of T-invariance.
In the case that neutrino obeys pure Bose statistics its equilibrium 
distribution is given by the standard Bose form (\ref{f-b-nu}). Indeed, 
it is easy to see that for this distribution $F$ vanishes and together
with it the collision integral vanishes too. 
However, the form of equilibrium distribution for mixed statistics is
not so evident. It depends upon an assumption about $F$
for particles obeying mixed statistics. We do not have rigorous arguments
in favor of one or other form for $F$ and as a reasonable guess assume
that the factor depending upon the neutrino statistics in $F$ changes as
\be
(1- f_\nu) \rar c^2 (1 - f_\nu) + s^2 (1 + f_\nu)
\nonumber
\label{1way}
\ee
where $c=\cos \gamma$, $s=\sin \gamma$ and $\gamma$ is some mixing angle
characterizing admixture of wrong statistics. 

Another possibility for description of mixed statistics in kinetic
equation could be
\be
(1- f_\nu) \rar  c^2 (1 - c^2 f_\nu) + s^2 (1 +c^2 f_\nu).
\nonumber
\label{2way}
\ee
However, it is easy to see that these two seemingly reasonable possibilities
(\ref{1way}) and (\ref{2way}) are identically equivalent. In both cases
\be
(1- f_\nu) \rar (1-\kappa f_\nu)
\label{st-mdf}
\ee
where
\be
\kappa = c^2 -s^2 = \cos 2\gamma
\label{kappa}
\ee
We call $\kappa$ the Fermi-Bose mixing parameter\cite{bbn}. 
One can check that in the case of mixed statistics introduced 
to kinetic equation according to (\ref{st-mdf})
the equilibrium distribution takes the form\cite{bbn}:
\be
f^{(eq)}_\nu = \left[ \exp (E/T) + \kappa \right]^{-1} \, .
\label{f-mixed}
\ee
where $\kappa$ runs from $+1$ to $-1$ corresponding respectively to 
Fermi and Bose limits. The
intermediate value $\kappa =0$ corresponds to Boltzmann statistics.

If $-1<\kappa <0$, the maximum value of the chemical potential may be
bigger than the neutrino mass:
\be 
\mu^{(max)} = m_\nu - T \ln (-\kappa)
\label{mu-max}
\ee
Bose condensation might take place for negative $\kappa$ only.

Another possible form of a modification of statistics dependent factor
in kinetic equation would emerge if we assume that there are two
neutrino fields with the same mass and different statistics, fermionic
and bosonic.  The Lagrangian would always depend upon two independent
field operators in the combination:
\be
\psi_\nu = c\psi_b + s \psi_f,
\ee
where $\psi_{f,b}$ are respectively bosonic and fermionic operators. 
In this case kinetic equation would contain two different terms:
\be
c^2 f_f (1- f_f)
\ee
and
\be
s^2 f_b (1 - f_b)
\ee
Equilibrium distributions would be canonical ones, e.g. for vanishing chemical
potential they are:
\be
f_{f,b} = 1/\left[\exp\left(E/T \right) \pm 1\right]
\label{ffb}
\ee
but the number of states in equilibrium becomes doubled. On the other 
hand, the
probability of e.g. neutron beta-decay remains the same as in the standard 
theory because $c^2 + s^2 = 1$.

\section{Big Bang Nucleosynthesis \label{s-bbn}}

The impact of purely bosonic neutrinos on BBN was considered earlier in
paper\cite{cgt}. The effects of mixed statistics described by the 
equilibrium distribution (\ref{f-mixed}) were calculated in our
work\cite{bbn}. The equilibrium energy density of bosonic 
neutrinos at $T\gg \mnu$ is
8/7 of the energy density of fermionic neutrinos and thus the
change
of statistics would lead to an increase of the effective number of
neutrino species at BBN by $\Delta N_\nu =3/7$ (for three neutrinos).
On  the other hand, a  larger
magnitude of the neutrino distribution function and the fact that it
enters kinetic equation (see (\ref{F}))  as $(1+f_\nu)$ instead of
$(1-f_\nu)$ makes
the weak reactions of neutron-proton transformations 
faster and hence the $n/p$ freezing temperature becomes lower.
This effect
dominates and as a result the effective number of massless species
becomes smaller than 3. According to the calculations of ref.\cite{bbn}
the effective number of neutrino species in the case of pure Bose
statistics becomes $N_{eff} = 2.43$, practically independently on the
value of the baryon-to-photon ratio $\eta =n_B/n_\gamma$.

\begin{figure}
\centering
\vspace*{13pt}
\leftline{\hfill\vbox{\hrule width 5cm height0.001pt}\hfill}
\mbox{\epsfig{figure=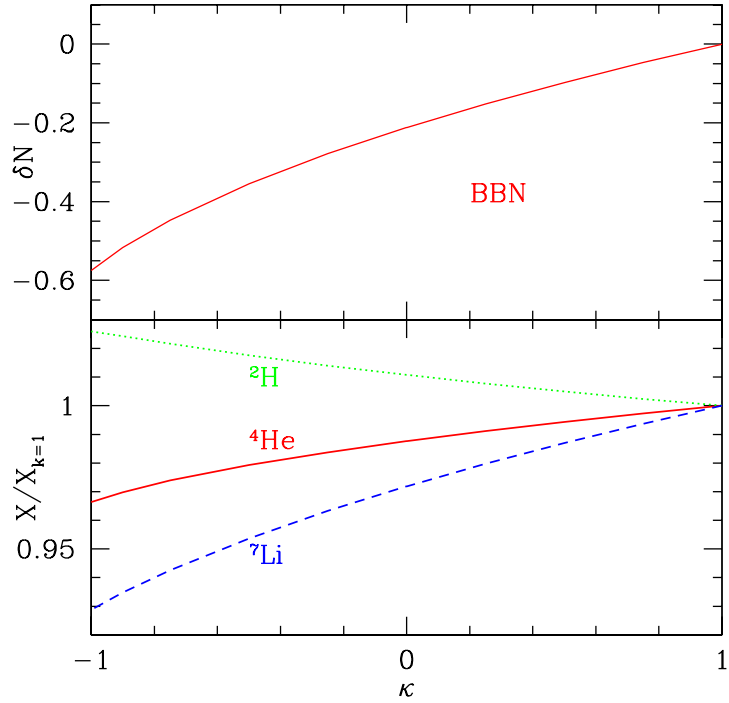,width=10.0cm}}
\vspace*{0.3truein}		
\leftline{\hfill\vbox{\hrule width 5cm height0.001pt}\hfill}
\caption{Upper panel: the change in the effective number of degrees of
freedom which corresponds to the change of the $^4$He abundance as a
function of the effective Fermi-Bose parameter $\kappa$.
Lower panel: the relative change of the
primordial abundances of deuterium, helium-4, and lithium-7, as
functions of $\kappa$.
We take $\eta = n_B/n_\gamma =6.5 \cdot 10^{-10} $.}
\label{fig-comb}
\end{figure}

The effective number of neutrino species determined by the comparison of
the calculated abundance of primordial $^4 He$ with the standard
result is presented in the upper panel of fig.~\ref{fig-comb}
as a function of $\kappa$. However, the
effect of change of statistics cannot be described by a simple change
in $N_\nu$ if other light elements are included.
In the lower panel of fig.~1 the relative changes of the abundances of
$^2 H$, $^4 He$, and $^7 Li$ are presented. As expected the mass
fraction of $^4 He$ drops down, while the amount of $^2 H$ goes up.  A
higher deuterium abundance can be explained by a slower conversion
of deuterium to heavier elements due to fewer neutrons and faster
cosmological expansion at $T\approx 0.8*10^9$ when the light elements
have been formed.

At $\kappa=-1$ we find
for $^4$He: $Y_p =0.240$, which makes much better agreement
with the value extracted from
observations (for a review of the latter see {\it e.g.}~\cite{bbn-obs}).
Different helium observations yield different results, {\it e.g.}, 
ref.~\cite{fieldsolive} finds $Y=0.238 \pm 0.002$, and ref.~\cite{izotovthuan}
finds  $Y = 0.2421 \pm 0.0021$ ($1\sigma$, only statistical error-bars).
These results are shown in figure 2 as the skew hatched (yellow) region.
Whether the existing helium observations are accurate or slightly
systematically shifted will be tested with future CMB
observations~\cite{trotta}.
\begin{figure}
\centering
\vspace*{13pt}
\leftline{\hfill\vbox{\hrule width 5cm height0.001pt}\hfill}
\mbox{\epsfig{figure=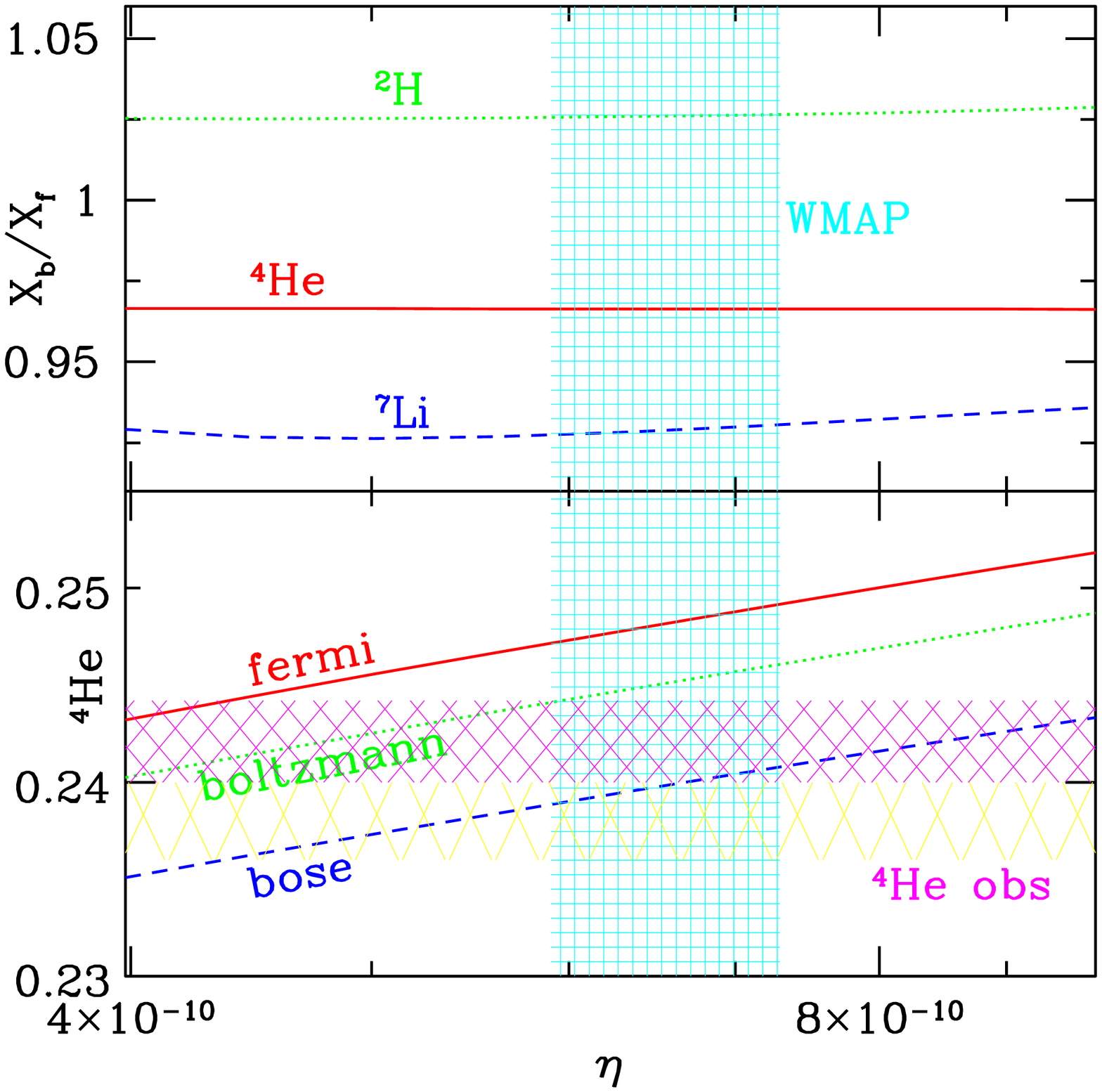,width=10.0cm}}
\vspace*{0.3truein}		
\leftline{\hfill\vbox{\hrule width 5cm height0.001pt}\hfill}
\caption{
Upper panel: the ratios of abundances
of different elements in the cases of purely
bosonic neutrinos with respect to the standard
fermionic case as functions of
the baryon number density, $\eta$. The vertically hatched (cyan) region
shows the WMAP $2\sigma$ determination of $\eta$. Lower panel: the
absolute abundance of $^4$He as a function of $\eta$ for the purely
bosonic, Boltzmann, and fermionic neutrino distributions, corresponding
to $\kappa=-1, 0, +1$ respectively. The two skew hatched
regions show the observation of primordial
helium from ref.$^{24)}$
(lower, yellow) and ref.$^{25)}$
(upper, magenta), which marginally
overlap at $1\sigma$.}
\label{fig-obs}
\end{figure}

The amount of $^2$H rises at most to $X_{^2H}/X_H = 2.5\cdot 10^{-5}$ and
the agreement between BBN and WMAP data remains good,
bearing in mind the observational uncertainties.
Primordial $^7$Li drops down to  $X_{^7 Li}/X_H = 4.55\cdot 10^{-10}$,
again slightly diminishing the disagreement between theory and observations.

We see that at the present time BBN does not exclude even a pure
bosonic nature of all three neutrinos. Furthermore, the agreement between
the value of the baryonic mass density, $\eta$, inferred from CMBR
and the predicted abundances of $^4$He, $^2$H, and $^7$Li  becomes
even better. In other words, in the standard BBN model
there is an indication of disagreement between observations
of $^4$He and $^2$H - they correspond to different values of $\eta$ with
the observed abundances of $^4$He indicating a smaller value than
given by CMBR, while $^2$H agrees with CMBR. Motivated by these results
the value of $\Delta N_\nu = -0.7\pm 0.35$ was suggested 
in ref.~\cite{steig}.
In the case of predominantly bosonic neutrinos, as discussed above, the
discrepancy between $^2$H, $^4$He, and CMBR disappears.

The results presented in this section are obtained for negligible
chemical potential of electronic neutrinos. On the other hand, formation
of cosmological neutrino condensate discussed in sec. 2
demands the maximum value of $\mu_\nu$ given by eq. (\ref{mu-max}). As is 
known\cite{lept-asym}, BBN allows at most $\mu/T = 0.07$ for any neutrino
flavor. This implies $\kappa > 0.9$. Such a large admixture of a wrong
bosonic state to $\nu_e$ is most probably excluded by the data on 
double beta decay (see below, sec. 7).
However, one may still hope
to save the neutrino cold dark matter ($\nu$CDM)
if the mixing angle determined from the decay
is different from that that enters neutrino kinetics (see discussion
in sec. 8.
Another possibility to save $\nu$CDM is to assume
that the chemical potentials of $\nu_{\mu}$ or $\mu_\tau$ are much larger 
than that of $\nu_e$. The latter are very weakly restricted by BBN and
only the large mixing between neutrino flavors equalizes all chemical
potentials. However, the change of neutrino statistics may lead
to a different refraction index in the primeval plasma and to suppression
of the transformation of $\nu_{\mu,\tau}$ into $\nu_e$. 
There is also a more conventional way to suppress 
neutrino flavor oscillations in the primeval plasma introducing
neutrino coupling to light pseudo-goldstone boson, Majoron. The effective
potential of neutrinos induced by the Majoron exchange would suppress
flavor transformations in the cosmological plasma\cite{ad-ft}. This would
allow to have large chemical potentials of $\nu_{\mu,\tau}$ and small
chemical potential of $\nu_e$.

\section{Astrophysical consequences \label{s-astro}}

Neutrino statistics plays key role in the environments where
neutrinos form dense degenerate gases.
Direct test of the ``bosonic'' nature of neutrinos can be provided
by precise measurements  of the neutrino energy  spectrum from supernova.
Generically, the spectrum  of bosonic neutrinos should be more narrow.
To establish the difference one  needs to measure
the spectrum both in the low,
$E < 3T$,  and in the high, $E > 3T$ energy parts.  

A violation of the exclusion principle can  influence
dynamics of the SN collapse. According to the usual scenario
at the initial stages (formation of the hot proto-neutron star)
the neutronization
leads to production of high concentration of the electron neutrinos
which are trapped in the core. The chemical potential of these neutrinos
(due to the Pauli principle) can reach 70 - 100 MeV. These neutrinos heat
the medium and diffuse from  the core. Violation of the Pauli principle 
allows for the neutronization neutrinos to be produced with lower energies.
These neutrinos escape easier the star leading to faster cooling
and lower central temperatures. The evolution of the lepton number would 
change as well.

High neutrino density in the condensate
(especially if an additional clustering occurs)
enhances the rate of the  $Z^0$-bursts produced by the annihilation 
of the ultra high energy (UHE)  cosmic
neutrinos on the relic neutrinos~\cite{z-burst}.
This in turn, enhances  production of the UHE
cosmic rays, and  may help to explain the cosmic ray evens above  the
GZK cut-off.

Charge asymmetric neutrino condensate may produce a strong refraction  of
the high energy neutrinos from remote sources (active galactic nuclei, gamma
ray bursters).  Apart from lensing, one may expect a substantial
impact on neutrino oscillations~\cite{LS}.
 
Since the density of dark matter in galaxies is about 6 orders of
magnitude larger than their average cosmological energy density,
a condensation of cold neutrinos around the Earth might have an
effect on the end point of the beta decay spectra,  in particular, in the
tritium decay experiments on search for neutrino mass~\cite{trit}.

\section{Double beta decay \label{s-2beta}}

In contrast to electrons and nucleons which form atoms and nuclei, where 
the effects of statistics are of primary importance, it is difficult to
observe processes with identical neutrinos. A realistic reaction for the
test of neutrino statistics can be the two-neutrino double beta decay, 
\be
A\rar A'+ 2\bar{\nu} + 2e^- 
\label{2-beta}
\ee
(or similar with production of antineutrinos and positrons). The  
probability of the decay as well as  the energy spectrum
and angular distribution of electrons should be affected by the change
of neutrino statistics.

To have a formalism for description of identical neutrinos one needs
to specify operators of neutrino creation/annihilation. We assume that 
they consist of two parts, fermionic, $\hf$, and bosonic $\hb$ for 
operators of annihilation, $\hat a = \hf + \hb $. Its  Hermitian 
conjugate could naturally be the operator of neutrino creation.   
Correspondingly we define one neutrino state as:
\be
|\nu\rangle =  \hat a^+ |0\rangle  \equiv
c_1 \hat f^+ |0\rangle + s_1 \hat b^+ |0\rangle
= c | f\rangle + s | b\rangle
\label{nu-state}
\ee
where $| f\rangle$ and $| b\rangle$ are respectively one particle
fermionic and bosonic states and 
$c_1=\cos \delta$ and {$s_1=\sin \delta$}. It would be natural to
expect that $\delta$ is equal to
$\gamma$ introduced above in eq. (\ref{1way}) 
but we cannot prove it formally.

To describe the two-neutrino state one needs to specify the relevant
commutators which, we postulate, have the following form:
\be
\hf \hb = e^{ i \phi} \hb \hf,\,\,\,
\hfc \hbc = e^{ i \phi} \hbc \hfc, \,\,\,
\hf \hbc = e^{ - i \phi} \hbc \hf,\,\,\,
\hfc \hb = e^{- i \phi} \hb \hfc,
\label{ab-af}
\ee
where $\phi$ is an arbitrary phase. The two neutrino state is natural
to define as
\be
|k_1,k_2\rangle = \hat a_1^+ \hat a_2^+ |0\rangle
\label{two-nu}
\ee

The matrix element of the decay of nucleus $A$ into $2\nu+2e+A'$ may be
possibly taken in the usual way: 
\be
A_{2\beta} = \langle k_1, k_2, 2e, A'\,{\Bigl\vert}  \int d^4 x_1 d^4 x_2
\psi_\nu (x_1) \nonumber\\
\psi_2 (x_2) {\cal M}(x_1,x_2) {\Bigl\vert} 0, A \rangle.
\label{A-2beta}
\ee
After making the necessary commutating according to  eq. (\ref{ab-af})
we obtain:
\be
A_{2\beta} = A_- \left[ c_1^4 + c_1^2 s_1^2 
\left( 1-\cos \phi \right)\right]
+ A_+ \left[ s_1^4 + c_1^2 s_1^2 \left( 1+\cos \phi \right)\right].
\label{A-2beta-2}
\ee
where $A_-$ and $A_+$ are respectively antisymmetric (fermionic) and
symmetric (bosonic) parts of two neutrino emission.
It is easy to see that the amplitude can be parametrized as
\be
A_{2\beta} =\cos^2 \chi\, A_- + \sin^2 \chi\, A_+,
\label{A-2beta-3}
\ee
where $\cos^2\chi =  c_1^4 + c_1^2 s_1^2 \left( 1-\cos \phi \right)$
and  $\sin^2\chi =  s_1^4 + c_1^2 s_1^2 \left( 1+\cos \phi \right)$.
The probability of the double beta decay integrated over neutrino 
momenta evidently does not contain interference between $A_+$ and
$A_-$ and is equal to:
\be
W_{tot} = \cos^4 \chi\, W_- + \sin^4 \chi\, W_+,
\label{W-tot}
\ee
where $W_{\pm}$ are proportional to $|A_\pm |^2$.

The probability of decay into unusual bosonic neutrinos is 
proportional to the bi-linear combinations of the type
$K_m K_n$,  $K_m L_n$, $L_m L_n$, where
\begin{eqnarray}
K_m^b \equiv [E_m  - E_i + E_{e1} + E_{\nu 1}]^{-1} - [E_m  - E_i + E_{e2} +
E_{\nu 2}]^{-1},
\nonumber\\
L_m^b \equiv [E_m  - E_i + E_{e2} + E_{\nu 1}]^{-1} - [E_m  - E_i + E_{e1} +
E_{\nu2}]^{-1}.
\label{prop}
\end{eqnarray}
Here the upper index $b$ indicates that the results are applicable
to bosonic neutrinos,
$E_i$ is the energy of the initial nuclei, $E_m$ is the energy
of the
intermediate nucleus state $m$, $E_{ei}$, and $E_{\nu i}$ are the energies
of electrons and neutrinos respectively.
The minus signs between the two terms in the above expressions are
due to the bosonic character of neutrinos;  in the case of  fermionic
neutrinos we would have
plus signs~\cite{boehm}. For electrons we assume the normal Fermi
statistics.

In the case of   $0^+ \rightarrow 0^+$ transitions the combinations  $K_m$
and $L_m$ can be approximated by
\be
K_m^b \approx \frac{E_{e2} - E_{e1} + E_{\nu2} - E_{\nu1}}{( E_m  - E_i
+ E_0/2 )^2},
~~~~~L_m^b \approx \frac{E_{e1} - E_{e2} + E_{\nu 2} - E_{\nu1}}{( E_m -
E_i + E_0/2 )^2},
\label{KL}
\ee
whereas for the  fermionic neutrinos
\be
K_m^f \approx L_m^f \approx  \frac{2}{E_m  - E_i + E_0/2 }.
\ee
Here $E_0/2 = E_e + E_{\nu}$ is the average energy of the leptonic pair.
Appearance of the differences of the electron and neutrino
energies in eq. (\ref{KL}) leads to a suppression of the total probability.
It also modifies the energy distributions of electrons.
The probabilities of the transitions $0^+ \rightarrow 2^+$
are proportional to the combinations $ (K_m - L_m)(K_n - L_n)$,
where
\be
(K_m^b - L_m^b) \approx  \frac{2(E_{e2} - E_{e1})}{(E_m  - E_i +
E_0/2)^2}.
\ee
In the case of fermionic neutrinos
the combination has an additional factor $(E_{\nu2} - E_{\nu1})/(E_m  -
E_i + E_0/2)$ and the suppression is stronger.

A simple estimate shows that the probability of $0^+ \rar 0^+$-transition
for bosonic neutrinos is suppressed by 1/250 for $^{56} Ge$ and
by 1/10 for $^{100} Mo$. Theoretically the total decay rate is known
with the accuracy within a factor of few. This probably allows to 
exclude a 100\% bosonic neutrino. However, the fraction of bosonic
neutrino can still be very high. According to our preliminary 
calculations\cite{ad-as,domin} the value of the mixing angle can
be as large as:
\be
\sin^2 \chi \leq 0.7
\label{2beta-lim}
\ee

For $0^+ \rar 2^+$ the situation is opposite: bosonic neutrinos are more
efficiently produced. However, no interesting bound is obtained 
in this case because
the statistics for these decays is much lower.

One can use the data on the spectrum of the emitted electrons, either 
single electron spectrum or distribution over the total energy of both
electrons. The spectra do not have any noticeable ambiguity related to 
unknown nuclear matrix elements and the present day accuracy is at the 
level of 10\%. Potentially their analysis may improve the above 
quoted limit (\ref{2beta-lim}) or indicate the existence 
of a ``bosonic''
admixture to neutrinos. Some already observed anomalies may be interpreted
as hints supporting the latter.

Unfortunately we cannot say at the present stage how the Fermi-Bose
parameter introduced above (\ref{kappa}) is related to 
the mixing angle $\chi$. Even if we assume that the mixing angle
in neutrino kinetics (\ref{1way}) is the same as in the definition
of neutrino states (\ref{nu-state}), the unknown value of the angle
$\phi$ which enters the commutation relations (\ref{ab-af})
and upon which depends the angle $\chi$ (\ref{A-2beta-3}) makes
the relation between $\kappa$ and $\chi$ rather arbitrary. 

\section{Theoretical problems and discussion \label{s-disc}}

Mentioned above ambiguities are related to intrinsic problems of
formulation a theory with mixed statistics. Working at a naive
level, as we did above, it is even difficult to define the properly
normalized particle number operator. According to eqs. (\ref{nu-state})
and (\ref{two-nu}) it is natural to define
the $n$ identical neutrino state as
\be
| n\rangle = \left( \cm f^+ + \sm b^+\right)^n |0\rangle
\label{n-part}
\ee
The normalization of this state is
\be
\langle n | n\rangle = \sm^{2(n-1)}\left[
n! \sm^{2} + (n-1)! \cm^2
\left(\frac{\sin \left(n\phi/2\right)}{\sin
\left(\phi/2\right)}\right)^2
\right]
\ee
If we introduce the particle number operator in the usual way:
\be
\hat n = a^+ a,
\ee
then its diagonal matrix elements would be
\bea
\langle n |\hat n | n\rangle = \sm^{2(n-1)}\,
\left[
n\,n! \sm^4 + 2n! \cm^2 \sm^2 \cos\frac{\phi (n-1)}{2}\,
\frac{\sin n\phi/2}{\sin \phi/2}\right. + \nonumber \\
\left. \cm^2 \left(n! \sm^2 + (n-1)! (\cm^2 - \sm^2)\right)
\left(\frac{\sin \left(n\phi/2\right)}{\sin
\left(\phi/2\right)}\right)^2
\right]
\label{n-n-n}
\eea
The particle number operator, as introduced above,
has reasonable and self-consistent interpretation
only for the case of pure statistics, while for mixed statistics it
even does not commute with Hamiltonian if the latter, or operator 
$\hat n$, or both are not somehow modified.

There are no problems with reactions where only one neutrino is
involved, but serious difficulties may arise with two neutrino 
reactions, as e.g. $\nu l \rar \nu l$ or $\bar \nu \nu \rar \bar l l$,
even if the participating neutrinos are not in an identical state. 
The amplitude of $\nu e$-elastic scattering in the usual approach is
given by the expansion of the T-exponent of the action and is
described by two diagrams differing by an interchange of emission
and absorption points. If taken literally, the diagrams with 
$W^\pm$-exchange would give vanishing amplitude for purely bosonic 
neutrinos
\footnote{This was noticed also by F. Vissani at this Conference.}.
In this case only $Z$-exchange would contribute to $\nu_e e$-scattering
and the cross-sections of $\nu_\mu e$ and $\nu_e e$-scattering 
would be equal. Reactor neutrino experiments are consistent with the
standard value of $\nu_e e$ cross-section and seem to exclude the
possibility of purely bosonic $\nu_e$. Using these data one can put a 
rather strong bound on bosonic admixture to electronic neutrino. On the
other hand, perturbation theory with non-bosonic Hamiltonian may need
to be modified and the above conclusion of vanishing of the amplitude
of scattering of pure bosonic neutrinos on electrons
would be invalid.

It is unclear if all these problems can be resolved in a simple way
or drastic modifications of the underlying theory is necessary, which
is a nontrivial task because the observed consequences of the theory
must not be destroyed.

The presentation in the previous sections 
and in the original paper\cite{ad-as} was
on pure (and poor) phenomenological level. For example if neutrinos
have mixed statistics then in double beta decay the symmetry of the
final state of neutrinos is mixed: symmetric with wight $a_+$ and
antisymmetric with the weight $a_-$. It seems plausible that these
weights are respectively $\cos^4 \chi$ and $\sin^4 \chi$ as argued
in the previous section, eq. (\ref{W-tot}), simply on the basis
on the normalization arguments.

Similar reasoning is possible for kinetic effects, 
eqs. (\ref{1way},{\ref{2way}).
There are no rigorous theoretical arguments in favor of such 
description but the result (\ref{f-mixed}) for the equilibrium
distribution in the case of mixed statistics
looks quite beautiful. Moreover, the fact 
that two ``reasonable'' (or natural) ways of description (\ref{1way})
and (\ref{2way}) give the same result is an argument in favor of
their validity.

\section{Conclusion}

There is no consistent theoretical frameworks for description of 
mixed neutrino statistics and even in the case of purely bosonic
neutrinos the fermionic property of the Hamiltonian would make a 
possible future theory quite unusual if it will ever be formulated.
Still independently on theory there could be some predictions
testable by experiment. So to summarize we will conclude that:
\begin{enumerate}
\item{}
The suggestion of bosonic or mixed statistics for particles (neutrinos)
with half integer spin looks exciting but opens a Pandora box of
serious theoretical problems, which may be impossible to resolve without
revolutionary modification of the standard theory. Such modification
looks especially difficult in the case of mixed statistics.
\item{}
The suggested mixture of statistics allows to break plenty of sacred
principles, as e.g. Lorentz invariance, CPT-theorem, locality, etc, 
which are actively discussed now.
\item{}
Bosonic neutrinos open a possibility of making all cosmological dark
matter out of neutrinos in accordance with Occam's razor: 
``Plurality should not be posited without necessity.''
\item{}
``Bosonization'' of neutrinos leads to effective number of neutrino 
species at BBN smaller than 3 and makes an agreement of the BBN 
calculations with the data noticeably better.
\item{}
Analysis of accumulated and accumulating data on two neutrino double
beta decay could restrict the admixture of wrong statistics to 
neutrinos or to indicate a violation of spin-statistics relation.
\item{}
Last, but not the least, if the validity of spin-statistics 
theorem has been studied for the usual matter, electrons and nucleons
it surely worth studying for neutrinos. The possibility that statistics
is modified for neutrinos seems more plausible because neutrino is
a natural particle to be a messenger from hidden sector of physics where
some principles respected in our world can be violated.
\end{enumerate}

\end{document}